\documentclass[sigconf,nonacm]{acmart}

\AtBeginDocument{%
  }

\usepackage{titlesec}

\definecolor{mygreen}{RGB}{0,150,0}
\definecolor{myred}{RGB}{255,0,0}
\usepackage{enumitem}

\usepackage[font=small,labelfont=bf]{caption}
\usepackage{array,ragged2e}
\newcolumntype{P}[1]{>{\RaggedRight\arraybackslash}p{#1}}

\definecolor{paired-light-blue}{RGB}{198, 219, 239}
\definecolor{paired-dark-blue}{RGB}{49, 130, 188}
\definecolor{paired-light-orange}{RGB}{251, 208, 162}
\definecolor{paired-dark-orange}{RGB}{230, 85, 12}
\definecolor{paired-light-green}{RGB}{199, 233, 193}
\definecolor{paired-dark-green}{RGB}{49, 163, 83}
\definecolor{paired-light-purple}{RGB}{218, 218, 235}
\definecolor{paired-dark-purple}{RGB}{117, 107, 176}
\definecolor{paired-light-gray}{RGB}{217, 217, 217}
\definecolor{paired-dark-gray}{RGB}{99, 99, 99}
\definecolor{paired-light-pink}{RGB}{222, 158, 214}
\definecolor{paired-dark-pink}{RGB}{123, 65, 115}
\definecolor{paired-light-red}{RGB}{231, 150, 156}
\definecolor{paired-dark-red}{RGB}{131, 60, 56}
\definecolor{paired-light-yellow}{RGB}{231, 204, 149}
\definecolor{paired-dark-yellow}{RGB}{141, 109, 49}

\definecolor{bg1}{HTML}{FF9966}
\definecolor{bg2}{HTML}{CCE5FF}
\definecolor{bg3}{HTML}{FFCC99}
\definecolor{bg4}{HTML}{FFC107}
\definecolor{bg5}{HTML}{FFCCCC}
\definecolor{bg6}{HTML}{D5E8D4}
\definecolor{bg7}{HTML}{eeeeee}
\definecolor{bg8}{HTML}{cdeb8b}
\definecolor{bg9}{HTML}{dae8fc}
\definecolor{bg10}{HTML}{a2e6eb}

\definecolor{bg31}{HTML}{FFCDD2} 

\definecolor{bg32}{HTML}{F8BBD0}

\definecolor{bg33}{HTML}{E1BEE7} 

\definecolor{bg34}{HTML}{D7CCC8} 

\definecolor{bg35}{HTML}{B2DFDB} 

\definecolor{bg36}{HTML}{A5D6A7} 

\definecolor{bg37}{HTML}{FFF9C4} 

\definecolor{bg38}{HTML}{FFECB3} 

\definecolor{bg111}{HTML}{CB6843}

\definecolor{bg112}{HTML}{D77C5C}

\definecolor{bg113}{HTML}{E28E6E}
\definecolor{bg114}{HTML}{E89F7D}
\definecolor{bg115}{HTML}{EDAE8A}
\definecolor{bg116}{HTML}{F0BA95}
\definecolor{bg117}{HTML}{F3C29F}
\definecolor{bg118}{HTML}{F6CCAA}
\definecolor{bg119}{HTML}{F8D5B3}
\definecolor{bg120}{HTML}{FADCBD}
\definecolor{bg121}{HTML}{FCE6C7}

\definecolor{bg39}{HTML}{FFE0B2} 

\definecolor{bg40}{HTML}{3CB371} 

\definecolor{bg43}{HTML}{ffe5d9}

\definecolor{bg15}{HTML}{7FFFD4}

\definecolor{bg17}{HTML}{F0FFFF}

\definecolor{bg18}{HTML}{F5FFFA}

\definecolor{bg19}{HTML}{F8F8FF}

\definecolor{bg20}{HTML}{FFFFFF}

\definecolor{bg21}{HTML}{E1F5FE}

\definecolor{bg22}{HTML}{B3E5FC}

\definecolor{bg23}{HTML}{81D4FA}

\definecolor{bg24}{HTML}{4FC3F7}

\definecolor{bg25}{HTML}{29B6F6}

\definecolor{bg26}{HTML}{03A9F4}
\definecolor{bg27}{HTML}{039BE5}
\definecolor{bg28}{HTML}{0288D1}
\definecolor{bg29}{HTML}{0277BD}
\definecolor{bg30}{HTML}{01579B}
\definecolor{bg16}{HTML}{FFCC99} 
\definecolor{pg51}{HTML}{E8F5E9} 
\definecolor{pg52}{HTML}{C8E6C9} 
\definecolor{pg53}{HTML}{B9F6CA} 
\definecolor{pg54}{HTML}{A9DFBF} 
\definecolor{pg55}{HTML}{BCF5A6} 

\definecolor{pg56}{HTML}{BEF1CE} 
\definecolor{pg57}{HTML}{CEF6EC} 
\definecolor{pg58}{HTML}{B7F0B1} 
\definecolor{pg59}{HTML}{B1F2B5} 
\definecolor{pg60}{HTML}{9DF3C4} 

\definecolor{pg61}{HTML}{DEF7E0} 
\definecolor{pg62}{HTML}{E8F8DC} 

\definecolor{pg63}{HTML}{EBF7E7} 
\definecolor{pg64}{HTML}{F0FDF4} 

\definecolor{pg65}{HTML}{F1FEE7} 
\definecolor{pg66}{HTML}{F7FFF6} 
\definecolor{pg67}{HTML}{FCFFE7} 
\definecolor{pg68}{HTML}{F4FFD2} 
\definecolor{pg69}{HTML}{EEFFE2} 
\definecolor{pg70}{HTML}{E3FDF5} 

\definecolor{connect-color}{RGB}{0,0,0}
\definecolor{middle-color}{RGB}{255,255,255}
\definecolor{leaf-color}{RGB}{173,216,230}
\definecolor{line-color}{RGB}{25,25,112}

\usepackage{url}            
\usepackage{booktabs}       
\usepackage{amsfonts}       
\usepackage{nicefrac}       
\usepackage{microtype}      
\usepackage{amsmath}
\usepackage{caption}

\usepackage{multirow}
\usepackage{subcaption}
\usepackage{listings}
\usepackage{tcolorbox}
\usepackage{xspace}
\usepackage{makecell}
\usepackage{soul}               
\setstcolor{red}            

\usepackage{longtable}
\usepackage{tablefootnote}

\usepackage[edges]{forest}
\definecolor{hidden-draw}{RGB}{20,68,106}
\definecolor{hidden-pink}{RGB}{255,245,247}
\definecolor{red}{RGB}{255,0,0}

\definecolor{hidden-draw}{RGB}{0,0,0}
\definecolor{hidden-pink}{RGB}{255,182,193}

\usepackage[T1]{fontenc}

\usepackage[utf8]{inputenc}

\usepackage[draft,textsize=footnotesize,textwidth=15mm]{todonotes}

\usepackage{forest}    

\usepackage{tikz}      

\usepackage{hyperref}  

\tikzset{
    root style/.style={
        draw,
        rounded corners,
        fill=blue!30, 
        align=center,
        font=\bfseries
    },
    child style/.style={
        draw,
        rounded corners,
        fill=green!30, 
        align=center,
        font=\bfseries
    },
    grandchild style/.style={
        draw,
        rounded corners,
        fill=red!30, 
        align=center,
        font=\bfseries
    }
}

\tikzset{
  my-box/.style={
    rectangle,
    draw=hidden-draw,
    rounded corners,
    text opacity=1,
    minimum height=1.5em,
    minimum width=40em,
    inner sep=2pt,
    align=center,
    line width=0.8pt,
  },
  leaf/.style={
    my-box,
    minimum height=1.5em,
    text=black,
    align=center,
    font=\normalsize,
    inner xsep=2pt,
    inner ysep=4pt,
    line width=0.8pt,
  }
}

\begin{document}
\setcopyright{none}

\title{Qualitative Insights Tool (QualIT): LLM Enhanced Topic Modeling}


\author{Satya Kapoor}
\affiliation{%
\institution{Amazon}
\city{Vancouver}
\state{British Columbia}
\country{Canada}}

\author{Alex Gil}
\affiliation{%
\institution{Amazon}
\city{Arlington}
\state{Virginia}
\country{USA}}

\author{Sreyoshi Bhaduri}
\affiliation{%
\institution{Amazon}
\city{Arlington}
\state{Virginia}
\country{USA}}

\author{Anshul Mittal}
\affiliation{%
\institution{Amazon}
\city{Arlington}
\state{Virginia}
\country{USA}}

\author{Rutu Mulkar}
\affiliation{%
\institution{Amazon}
\city{Seattle}
\state{Washington}
\country{USA}}



\begin{abstract}

Topic modeling is a widely used technique for uncovering thematic structures from large text corpora. However, most topic modeling approaches e.g. Latent Dirichlet Allocation (LDA) struggle to capture nuanced semantics and contextual understanding required to accurately model complex narratives. Recent advancements in this area include methods like BERTopic, which have demonstrated significantly improved topic coherence and thus established a new standard for benchmarking. In this paper, we present a novel approach, the Qualitative Insights Tool (QualIT) that integrates large language models (LLMs) with existing clustering-based topic modeling approaches. Our method leverages the deep contextual understanding and powerful language generation capabilities of LLMs to enrich the topic modeling process using clustering. We evaluate our approach on a large corpus of news articles and demonstrate substantial improvements in topic coherence and topic diversity compared to baseline topic modeling techniques. On the 20 ground-truth topics, our method shows 70\% topic coherence (vs 65\% \& 57\% benchmarks) and 95.5\% topic diversity (vs 85\% \& 72\% benchmarks). Our findings suggest that the integration of LLMs can unlock new opportunities for topic modeling of dynamic and complex text data, as is common in talent management research contexts.

\end{abstract}


\maketitle


\begin{figure*}
   \centering
    \resizebox{\textwidth}{!}{%
    \includegraphics{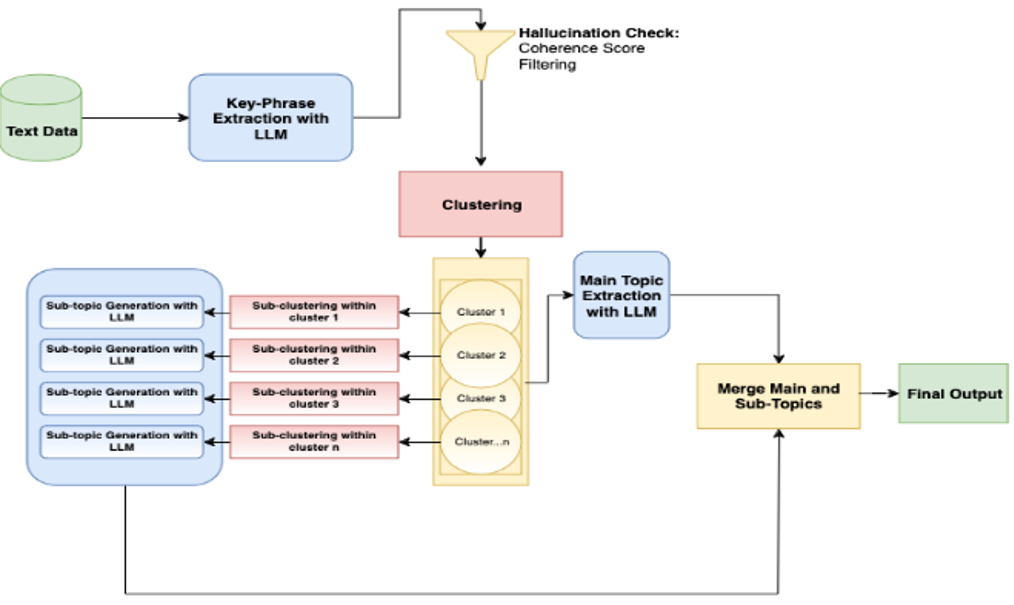}}
    \caption{QualIT : Qualitative Insights Tool}
    \label{fig:QualIT}
\end{figure*}

\section{Introduction}
Topic modeling is a widely-used natural language processing (NLP) technique for extracting latent thematic structures from unstructured text data, such as social media posts, news articles, or customer feedback \cite{blei2012probabilistic}\cite{grimmer2013text}. By employing probabilistic models to systematically identify and categorize patterns within the text, topic modeling enables researchers to uncover themes and perspectives that may not be immediately apparent to human analysts \cite{blei2003latent}. The flexibility of topic modeling allows it to be applied across a range of theoretical and epistemological frameworks, making it a valuable tool in both quantitative and qualitative research \cite{roberts2014structural}. 

Traditional topic modeling techniques (e.g. Latent Dirichlet Allocation) suffer from several limitations (e.g. bag-of-words limitation, specifying number of clusters) when compared to existing deep learning-based methods. They also fail to capture the contextual nuances and ambiguities inherent in natural language, as they rely heavily on predefined rules and patterns \cite{devlin2018bert}\cite{radford2019language}. This can make it challenging to handle the complexities and variations present in real-world text data, and may require domain-specific knowledge or fine-tuning to achieve acceptable performance \cite{lee2019patentbert}. Recent advancements in LLMs have positioned methods like BERTopic as a strong alternative to traditional topic modeling methods, and have largely overcome the limitations by leveraging deep neural networks to learn rich, contextual representations from large amounts of text data \cite{ashish2017attention}\cite{devlin2018bert}. These powerful models can capture subtle semantic and pragmatic features of language, and demonstrate strong generalization capabilities through transfer learning \cite{radford2019language}\cite{brown2020language}. They arent without limitations though. BERTopic, for example, is a clustering based approach which suffers from limitations such as word representation overload or generation of only one topic per text. 

In this paper, we present - QualIT : the Qualitative Insights Tool (pronounced "kwaa-luh-tee") to extend the capabilities of existing topic models. Our novel approach integrates pre-trained LLMs with clustering techniques to systematically address the limitations of both methods and generate more nuanced and interpretable topic representations from free-text data. Combining LLMs and clustering techniques can provide a powerful and scalable approach to automatically identify themes and patterns in large volumes of unstructured text data. LLMs offer some semantic understanding and the ability to capture contextual nuances \cite{maatouk2024large}, while clustering algorithms enable unsupervised grouping of similar responses into topics. Overall, the synergy between LLMs' natural language understanding and the clustering approach's ability to organize and summarize data can revolutionize topic modeling, providing a robust and insightful approach to analyzing text responses at scale.


\section {Topic Modeling in Talent Management Research} 
Talent management researchers leverage both psychological and data/applied science to provide actionable insights to managers, leaders, and HR professionals at an organization \cite{bhaduri2024reconciling}\cite{bhaduri2023using}. Voice of Customer (VOC) is an example of a key mechanism used by Talent Management researchers to collect feedback from customers and allows researchers in talent management to close the loop. VOC via surveys offer both qualitative and quantitative response options, both sources of important information for research teams to understand how customers interact with products AND what customers expect from products \cite{bhaduri2024multi}. 

However, insights from qualitative text largely remains a missed opportunity due to being time, labor, and resource intensive to analyze manually \cite{bhaduri2018nlp}. Typically, qualitative research projects take three months, end to end for a team of researchers, including sampling participants, collecting data, and analyzing documents. An automated topic modeling tool, such as QualIT, compliments the analysis step which is estimated to take one month of a researcher's time per project. In comparison, a pre-trained LLM model may take ~1:30 minutes to process 2500 documents. Of course, human in the loop deep-dives and quality checks would still be necessitated, however, such automated quantative approaches can provide research direction and support for qualitative researchers.

Further, the lack of familiarization/expertise in qualitative research methods limits easy analysis or sharing of insights. For program or product owners, topic modeling tools can democratize access to insights by automating and augmenting analysis of qualitative documents and thematically distilling the information. For researchers, these tools do not aim to replace manual deep-dives, but rather serves as a novice qualitative research assistant \cite{bhaduri2024reconciling} \cite{bhaduri2021semester} and reduces the amount of time it takes to manually analyze open text documents and surfaces effective trends in the data. 


\section{Related Work}
One of the most widely-used topic modeling approaches is Latent Dirichlet Allocation (LDA) \cite{blei2003latent}. LDA is a generative probabilistic model that operates on the principle that each document in a corpus is composed of a mixture of latent topics, with each topic being represented by a unique probability distribution over the vocabulary \cite{blei2012probabilistic}. The model learns these topic-word distributions by leveraging the co-occurrence patterns of words within the documents, allowing it to uncover the underlying thematic structure of the corpus.

A key step in the LDA modeling process is determining the appropriate number of topics to be extracted from the data. This number is not something LDA can automatically infer, but rather must be provided by the researcher as an input parameter \cite{griffiths2004finding}. The choice of the number of topics can have a significant impact on the interpretability and performance of the LDA model, as too few topics may fail to capture the nuances of the data, while too many topics can lead to overfitting and poor generalization \cite{arun2010finding}. Unfortunately, there is no universally optimal approach for selecting the number of topics, and the appropriate choice often depends on the specific research objectives and characteristics of the text corpus \cite{deveaud2014accurate}. As a result, practitioners must carefully explore and validate different topic count configurations to arrive at model parameters that best suits their needs.

The limitations of LDA have motivated researchers to explore the use of more advanced natural language processing methods, particularly those based on LLMs. LLMs, such as BERT \cite{devlin2018bert}, Generative Pre-trained Transformers (GPT) \cite{radford2019language}, and T5 \cite{colin2020exploring}, have demonstrated remarkable performance on a wide range of NLP tasks by leveraging deep neural networks to learn rich, contextual representations from massive amounts of text data. Unlike traditional topic modeling approaches that rely on hand-crafted features or simple statistical patterns, LLMs can capture complex semantic and syntactic relationships within language, allowing them to better handle the nuances and ambiguities inherent in real-world text \cite{liu2019roberta}. Moreover, these models can be fine-tuned on domain-specific data or downstream tasks, enabling them to adapt to the particular characteristics of a given text corpus \cite{sun2020ernie}. 

Recent studies have explored the integration of LLMs into topic modeling frameworks, demonstrating significant improvements in performance and interpretability compared to traditional methods. For example, some authors \cite{bianchi2020pre} have proposed a topic modeling approach that uses BERT embeddings to initialize the topic-word distributions, leading to more coherent and semantically meaningful topics. Other researchers have investigated the use of LLMs for various aspects of the topic modeling pipeline, such as document representation \cite{zhao2021contrastive}, and topic labeling \cite{hoyle2021automated}. These advancements have shown that integration of LLMs can make a more powerful topic modeling tool for uncovering insights from large text corpora.

\section{Methodology}
We propose a new method LLM Enhanced Topic Modeling, which consists of multiple steps to generate the main topics, which are then used towards determining sub-topics from documents. The three key steps in this approach are Key Phrase Extraction, Hallucination Check, and Clustering.

\subsection{Key-Phrase Extraction}
In this step, we prompt the LLM to extract key-phrases representing the individual document. The LLM analyzes the content, discerning patterns and topics that frequently occur within the text. Guided by the prompt, the LLM pinpoints the key-phrases related to the defined role. The LLM prompt can extract multiple key phrases from the document, depending on its content. These key-phrases are essential for understanding the more nuanced aspects of the main subject matter. Adding this step provides an advantage over alternative methods. Alternative methods (e.g.BERTopic) assume that each document only contain a single topic, when in reality a document may contain more than a single topic.

\subsection{Hallucination Check}
To ensure the reliability of extracted key-phrases, a coherence score is calculated for each phrase. This score assesses how well the key-phrase aligns with the actual text, serving as a metric for consistency and relevance of the subsequent topics. Key-phrases with the lowest coherence scores may be flagged for 'hallucination', indicating potential errors in topic extraction, and are removed accordingly. For our approach, phrases with coherence scores below 10\% were excluded. The coherence score was calculated using cosine similarity (Equation 1), with the texts first converted to embeddings using the Amazon Titan model.

\begin{equation}
\label{eq:2}
    C_i = \frac{1}{n} \sum_{j=1}^n \frac{\left(V_{\text{input},ij} \cdot V_{\text{keyphrases},ij}\right)}{\left|V_{\text{input},ij}\right| \cdot \left|V_{\text{keyphrases},ij}\right|}
\end{equation}
Where:
\begin{itemize}
    \item $C_i$ is the coherence score for the $i$-th document.
    \item $n$ is the number of dimensions in the embedding space.
    \item $V_{\text{input},ij}$ is the $j$-th dimension of the normalized embedding vector for the input text of the $i$-th document.
    \item $V_{\text{keyphrases},ij}$ is the $j$-th dimension of the normalized embedding vector for the theme text of the $i$-th document.
    \item $\left(V_{\text{input},ij} \cdot V_{\text{keyphrases},ij}\right)$ denotes the dot product of the two vectors.
    \item $\left\|v\right\|$ denotes the Euclidean norm (or length) of vector $v$.
\end{itemize}

\subsection{Clustering}
We use a partitional clustering algorithm (K-Means) to group the key-phrases identified in the previous step. The aim of this step is to group documents into multiple clusters, each representing a collection of documents with similar semantic content. In our approach, we implement two layers of clustering to find main topics and sub-topics. Unlike existing methods that directly use documents to create clusters, this approach leverages key phrases as they represent a more condensed form of the documents.

\subsubsection{Clustering for Main Topics}
Create the primary cluster and pass the key-phrases of each cluster to another LLM prompt to distill a main theme from the grouped documents. This synthesis involves extracting a comprehensive topic that encapsulates the essence of each cluster, providing a clear and consolidated view of the overarching subjects within the documents.

\subsubsection{Clustering for Sub Topics}
Implement a secondary level of clustering within each primary cluster to uncover more specific sub-topics. This will involve reapplying the clustering algorithm to each main cluster, separating the documents into sub-clusters based on finer nuances and more detailed thematic differences. For each sub-cluster, we prompt the LLM again to analyze condensed documents within sub-clusters. The model extracts sub-topics, for each sub-cluster. 

There are several advantages of using this novel approach for LLM extracted key-phrases as features for clustering, as opposed to direct grouping of documents. Primarily, it reduces noise and the influence of irrelevant data, allowing the algorithm to operate on the distilled thematic essence of the documents. Another key benefit is that this approach can avoid the need for manual data exploration and cleaning steps, such as stopwords or punctuation removal, as the LLM is able to extract only the most meaningful content from the documents. This results in a set of clusters that are thematically concentrated, facilitating a more nuanced analysis and understanding of the document collection. Our approach ensures that documents with shared underlying topics are clustered together, even if they are not textually identical. One major drawback of K-Means is that it requires the number of clusters as a parameter to perform clustering. To address this drawback, we utilized the length of data and calculate a Silhouette score (Equation 2) to automatically determine the number of clusters. Silhouette score is a metric used to calculate the ideal number of clusters \cite{shahapure2020cluster}.

\begin{equation}
\label{eq:2}
    s(i) = \begin{cases}
        \frac{b(i) - a(i)}{\max{a(i), b(i)}}, & \text{if } |C_I| > 1 \\
        0, & \text{if } |C_I| = 1
    \end{cases}
\end{equation}

Where:
\begin{itemize}
    \item s(i): Silhouette score
    \item a: The mean distance between a sample and all other points in the same cluster.
     \item b: The mean distance between a sample and all other points in the nearest cluster that the sample is not a part of.
\end{itemize}

\begin{table*}[h!]
    \begin{tabular}{cccc}
        \hline
        comp.graphics & talk.politics.guns & rec.sport.baseball & sci.crypt \\
        comp.os.ms-windows.misc & talk.politics.mideast & rec.sport.hockey & sci.electronics 
        \\
        comp.sys.ibm.pc.hardware & misc.forsale & rec.autos & sci.med 
        \\
        comp.sys.mac.hardware & talk.politics.misc & alt.atheism & sci.space 
        \\
        comp.windows.x & talk.religion.misc & soc.religion.christian & rec.motorcycles \\
        \hline
        \\
    \end{tabular}
    \caption{Ground-truth topics from 20 NewsGroup}
    \label{tab:table-label}
\end{table*}

\begin{table*}[h!]
    \begin{tabular}{cc}
        & List of Words \\
        \hline
        Output One & ['sale', 'discount', 'price', 'pricing', 'prices', 'purchase', 'dealer', 'sales', 'offer', 'shipping'] \\
        \hline
        Output Two & ['space', 'launch', 'spacecraft', 'lunar', 'nasa', 'satellite', 'orbit', 'rocket', 'moon', 'satellites'] \\
        \hline
        Output Three & ['encryption', 'security', 'cryptography', 'key', 'privacy', 'crypto', 'decryption', 'data', 'secure', 'vulnerabilities'] \\
        \hline
        \\
    \end{tabular}
    \caption{Example output from QualIT}
    \label{tab:table-label}
\end{table*}

\section{Experimental Setup}
AWS Sagemaker and AWS Bedrock were used to preprocess data, run experiments and validate results. For the benchmark models, Gensim’s LDA implementation and BERTopic were used.

\subsection{Dataset}
The 20 NewsGroups  dataset  was used to run the experiments. This dataset was chosen as a benchmark because it is widely used for these type of experiments with a recent example in the BERTopic paper. The 20 NewsGroups dataset3 contains 20,000 news articles across 20 categories. The dataset was pre-processed minimally: lowered tokens, special characters, stop words and tokens smaller than length 3 were removed, as well as tokens were lemmatized. This pre-processed dataset was used in all methods in this experiment for fair comparisons.

\subsection{Model}
LLM Enhanced Topic Modeling will be compared with LDA and BERTopic. Both LDA and BERTopic ran with default parameters from their respective model providers. Our LLM Enhanced Topic Modeling utilized Claude-2.1, with a temperature setting of 0, top\_k of 50, and top\_p of 0. These parameters were chosen with the aim of achieving the highest coherence scores, reflecting its superior semantic understanding and contextual analysis at the individual document level.

\subsection{Evaluation}

\begin{equation}
\label{eq:3}
i_n(x,y) = \frac{\left(\ln\frac{p(x,y)}{p(x)p(y)}\right)}{-\ln p(x,y)}
\end{equation}

We use topic coherence and topic diversity to measure performance of this experiment. We chose these metrics as they allow for comparisons between our benchmark models. Topic coherence (TC) (Equation 3) is a measure used to evaluate how meaningful a topic is based on the degree of semantic similarity among the top most frequent words within the topic. It ranges from [-1, 1] and is estimated by normalized pointwise mutual information \cite{bouma2009normalized}. A score of 1 in topic coherence indicates a perfect association. The importance of this metric is its semblance to human judgement with reasonable performance \cite{lau2014machine}. Topic diversity (TD) is the percentage of distinct words for all topics produced. \cite{dieng2020topic} This metric ranges from [0, 1], with 1 signaling diversity in topics. Each model was evaluated by using TC \& TD at intervals of 10 topics, within the search space from 10 to 50 topics, for a total of 5 evaluations. The results were averaged across all runs.

Table 1 is an example of the ground-truth topics. Table 2 is an example of output topic words by our proposed method. Each evaluated method outputs a similar list of words. The words in Table 2 are used to describe and understand what each output is meant to be about. We evaluate these outputs on how well they map to the 20 ground-truth categories in Table 1. This step is performed by manual human classification of each method output to the 20 ground-truth mapping. This step is important because it signals how well each method is able to correctly cluster to the ground truth and how easy it is for humans to classify and agree on the output’s classification. For quantifying how well the methods do, we calculate the percentage of agreement between evaluators on categorized topics. 

\begin{table*}[h!]
    \centering
    \begin{tabular}{cccc}
        \\ \
         & No. of Topics & Topic Coherence & Topic Diversity \\
        \hline
        & 10 & 47.0\% & 69.0\% \\
        & 20 & 57.0\% & 72.0\% \\
        LDA & 30 & 52.0\% & 67.3\% \\
        & 40 & 52.0\% & 79.1\% \\
        & 50 & 49.0\% & 77.0\% \\
        & Avg & 51.4\% & 72.7\% \\
        \\
        \hline
        & 10 & 56.0\% & 82.0\% \\
        & 20 & 65.0\% & 85.0\% \\
        BERTopic & 30 & 62.0\% & 88.3\% \\
        & 40 & \textbf{62.0}\% & 88.8\% \\
        & 50 & \textbf{60.2}\% & 87.2\% \\
        & Avg & 61.0\% & 86.3\% \\
        \\
        \hline
        & 10 & \textbf{66.0}\% & \textbf{95.0}\% \\
        & 20 & \textbf{70.0}\% & \textbf{95.5}\% \\
        \textbf{QualIT} & 30 & \textbf{65.0}\% & \textbf{93.0}\% \\
        & 40 & 61.0\% & \textbf{93.0}\% \\
        & 50 & 60.0\% & \textbf{92.0}\% \\
        & Avg & \textbf{64.4}\% & \textbf{93.7}\% \\  
        \\
        \hline
        \\
    \end{tabular}
    \caption{Each model was evaluated by using Topic Coherence \& Topic Diversity at intervals of 10 topics, within the search space from 10 to 50 topics, for a total of 5 evaluations. The results were averaged across all runs.}
    \label{tab:table-label}
\end{table*}

\begin{table*}[h!]
    \begin{tabular}{cccc}
        & & Percentage of Agreement for Categorized Topics &\\
        & At least 2 evaluators agreed & At least 3 evaluators agreed & All 4 evaluators agreed\\
        \hline
        LDA & 50\% & 25\% & 20\% \\
        BERTopic & 45\% & 25\% & 20\% \\
        QualIT & 80\% & 50\% & 35\% \\
        \hline
        \\
    \end{tabular}
    \caption{Agreement of topic classification by human evaluators to ground truth. A higher value indicates greater classification overlap between independent evaluators.}
    \label{tab:table-label}
\end{table*}

\section{Results}
The results from this experiment can be found in Table 3 and human evaluation of outputs can be found in Table 4. Table 3 shows that on average, LLM Enhanced Topic Modeling outperforms both LDA and BERTopic in the 20 NewsGroups dataset in terms of TC and TD. The proposed method exhibits greater performance for both Topic Coherence and Topic Diversity for topics in the 10-30 range. The 10-30 topic range is ideal for this dataset because 20 topics is the ground truth number for this dataset. All three methods achieved their highest level of coherence and diversity when the number of topics was set to 20. This finding is consistent with the structure of the dataset. However, when the number of topics increases to 40 and 50, BERTopic minimally outperforms LLM Enhanced Topic Modeling in terms of Topic Coherence, but not for Topic Diversity. 

Further, human evaluators were tasked with categorizing topic words (outputs) from each method into one of the 20 ground-truth categories in Table 1. We found that human evaluators were able to agree in topic classification more often with the outputs of QuallT vs the benchmarks, as shown in Table 4. For example, the column “At least 3 evaluators” shows us how many lists of topic words (out of 20) were identically mapped to the same topics by at least 3 evaluators independently. We see that the percentage of agreement for categorized topics is consistent across the number of evaluators and across methods. This means that based on our human evaluation, the output from our method is less ambiguous to humans and easier to classify into topics vs the benchmarks.

\section{Limitations and Future Work}
The current method presents certain limitations that could be addressed in future research. Firstly, the runtime of the model is a significant constraint for large dataset; efforts should be made to reduce it to be more in line with BERTopic, which completes in approximately 30 minutes as opposed to current 2-3 hours for our LLM Enhanced Topic Modeling. This enhancement in efficiency would greatly benefit scalability and usability in practical applications. Additionally, while our method currently utilizes K-Means clustering, there is an opportunity to explore HDBSCAN. Preliminary comparisons suggest that BERTopic’s use of HDBSCAN demonstrates more effective results than K-Means. Adding a similar approach may enhance the model’s ability to discern and group more complex and nuanced data patterns, thus offering a promising avenue for advancing the robustness and accuracy of the topic modelling performed by our method.

\section{Closing Thoughts}
Qualitative Inights Tool (QualIT) presented in this paper represents a significant advancement in the extraction and analysis of qualitative insights from unstructured text data. By leveraging state-of-the-art pre-trained large language models and a novel topic modeling framework, QualIT is able to surface both high-level topic insights as well as more granular subtopics from qualitative feedback data. Crucially, our experiments demonstrate that QualIT's approach produces more coherent and diverse topics compared to benchmark topic modeling techniques. 

As organizations increasingly rely on qualitative data to drive strategic decision making especially for talent management, tools like QualIT that can efficiently and effectively extract meaningful insights will become increasingly invaluable. We believe the QualIT framework represents an important step forward in empowering researchers, product teams, and decision-makers to uncover the rich insights hidden within their qualitative datasets. Going forward, further enhancements to the language modeling capabilities (such as languages beyond English, especially low resources ones) and topic clustering algorithms underpinning QualIT hold promise to unlock even more powerful qualitative analysis capabilities.

\bibliographystyle{ACM-Reference-Format}
\bibliography{sample-base}

\end{document}